# Key Gene Mining in Transcriptional Regulation for Specific Biological Processes with Small Sample Sizes Using Multi-network pipeline Transformer

Kerui Huang[1], Jianhong Tian[2], Lei Sun[3], Li Zeng[1], Peng Xie[1], Aihua Deng[1], Ping Mo[1], Zhibo Zhou[1], Ming Jiang[1], Yun Wang[1], Xiaocheng Jiang[2]

[1]Key Laboratory of Agricultural Products Processing and Food Safety in Hunan Higher Education, Science and Technology Innovation Team for Efficient Agricultural Production and Deep Processing at General University in Hunan Province, Hunan University of Arts and Science, Changde 415000, China.

[2]College of Life Sciences, Hunan Normal University, Changsha 410081, China.

[3]Key Laboratory of Research and Utilization of Ethnomedicinal Plant Resources of Hunan Province, College of Biological and Food Engineering, Huaihua University, Huaihua 418000, China.

## Abstract

Gene mining is an important topic in the field of life sciences, but traditional machine learning methods cannot consider the regulatory relationships between genes. Deep learning methods perform poorly in small sample sizes. This study proposed a deep learning method, called TransGeneSelector, that can mine critical regulatory genes involved in certain life processes using a small-sample transcriptome dataset. The method combines a WGAN-GP data augmentation network, a sample filtering network, and a Transformer classifier network, which successfully classified the state (germinating or dry seeds) of *Arabidopsis thaliana* seed in a dataset of 79 samples, showing performance comparable to that of Random Forests. Further, through the use of SHapley Additive exPlanations method, TransGeneSelector successfully mined genes involved in seed germination. Through the construction of gene regulatory networks and the enrichment analysis of KEGG, as well as RT-qPCR quantitative analysis, it was confirmed that these genes are at a more upstream regulatory level than those Random Forests mined, and the top 11 genes that were uniquely mined by TransGeneSelector were found to be related to the KAI2 signaling pathway, which is of great regulatory importance for germination-related genes. This study provides a practical tool for life science researchers to mine key genes from transcriptome data.

## Introduction

Gene mining is a series of methods that can reveal genes that play critical roles in specific biological processes,which plays a significant role in the life sciences. In plants, many important agronomic traits, such as yield and disease resistance, are complex quantitative traits that are regulated by multiple genes and environmental interactions[1-3], and to improve these traits, it is necessary to first identify their genetic mechanisms and key regulatory genes. In medical research, many diseases have genetic foundations, and an individual's susceptibility to a disease is related to gene variations and gene expression patterns[4]. Mining disease-related genes and revealing their roles in related cellular processes and signaling pathways can help understand the disease pathogenesis[4], identify disease-related biomarkers[5], and even be used to further identify therapeutic targets[6].

Currently, next-generation sequencing technology has generated a large amount of genetic data, providing great opportunities for gene mining. Among these technologies, transcriptome sequencing is a high-throughput, high-sensitivity transcriptome expression monitoring technology that can generate a large amount of expression data at the transcriptional level. Machine learning methods can effectively extract key genes related to specific phenotypes or diseases as biomarkers from high-dimensional transcriptome data, which is currently a popular key gene mining method[7]. Compared with traditional transcriptional data mining methods using fold change of gene expression levels to identify differentially expressed genes, the integration of transcriptional data and machine learning algorithms can achieve more accurate, comprehensive, and effective gene mining and functional prediction. For example, Li et al. used machine learning methods to extract key genes from gene expression data of

COVID-19 patients to evaluate disease severity[8]. Yu et al. applied machine learning and transcriptome sequencing to screen 9 SNPs in the transcriptome data of *Platycodon grandiflorus*, which can be used to identify flower color[9]. Pal et al. developed a tool based on support vector machines to predict disease-resistance proteins in plants[10]. Chen et al. developed a series of machine learning methods combinations to identify key genes in bovine multi-tissue transcriptome data to predict feed efficiency[11]. Although the combination of transcriptome sequencing and machine learning has performed well in key gene mining, traditional machine learning algorithms for gene mining usually require manual feature engineering to improve the performance, which results in the discard of many seemingly unimportant genes. However, genes in organisms exist in complex regulatory relationships and hidden action patterns[12,13], thus those discarded seemingly unimportant genes may be of key value. Given the limitations of traditional machine learning in gene mining, developing a machine learning algorithm that can mine key genes from transcriptomics data and fully capture the global interactions between genes is meaningful.

Deep learning is a subfield of machine learning that involves the use of neural networks to model and solve complex problems. It is based on artificial neural networks (ANNs), also known as deep neural networks (DNNs), which are inspired by the structure and function of the human brain's biological neurons. These networks are designed to learn from large amounts of data and can capture complex and nonlinear relationships within the data to solve complex machine learning problems [14-17]. Deep learning computation can capture complex nonlinear relationships, process unstructured data, have excellent prediction performance, and can automatically learn effective features from high-dimensional data without the need for manual feature selection, thus enabling a comprehensive understanding of the effectiveness of features on the global validity of the mode[18,19], thus can be used to fully mine key genes in life activities without feature selection. In addition, natural language processing (NLP) models in deep learning can capture long-distance dependencies between sequences, for example, RNN [20,21] and LSTM [22] can capture long-distance relationships in sequences, and are widely used in natural language processing tasks. The characteristic of capturing long-distance dependencies of NLP models can applied to capture complex regulatory relationships between genes, making it a highly promising machine learning algorithm for key gene mining. The Transformer model is a deep learning architecture introduced in 2017, primarily designed for NLP tasks, such as machine translation and time series prediction. It is based on the encoder-decoder architecture and relies only on the attention mechanism, which allows it to capture long-range dependencies and relationships between sequential elements more effectively than traditional recurrent neural networks (RNNs) and LSTMs [23] and has revolutionized the field of NLP[24-26]. Therefore, using natural language processing architectures in deep learning, especially Transformers, for classification tasks related to certain biological processes and mining key genes is a promising approach with potential applications.

So far, studies combining transcriptomic gene expression data with Transformer architectures have mainly focused on cell type classification tasks in single-cell sequencing rather than normal RNA sequencing. For example, TOSICA is a multi-head self-attention deep learning model based on Transformer that enables interpretable cell type annotation [27]; STGRNS is an interpretable transformer-based method for inferring gene regulatory networks from single-cell transcriptomic data [28]. Although these studies have explored a series of biological problems in single-cell sequencing, they have not addressed the problem of key gene mining, and the cost of single-cell sequencing is currently relatively high, and most biological processes are still studied using traditional transcriptomics sequencing methods. Therefore, developing a method for key gene mining using Transformer architectures combined with traditional transcriptomics sequencing data has strong practical significance. Currently, there are only two studies applying Transformer architectures to normal RNA sequencing data: T-GEM[29] is an interpretable deep learning model for cancer phenotype prediction based on gene expression data Another study is DeepGene Transformer[30], which is used for gene expression-based classification of cancer types. However, both of the two studies are not focused on key gene mining methods. Therefore, developing a method for key gene mining using Transformer models and transcriptomics data is necessary.

One of the challenges in combining standard transcriptomics sequencing data with deep learning is that the sample size of standard transcriptomics sequencing data in specific biological processes is relatively limited, making it difficult to meet the high sample size requirements of deep learning, which typically requires hundreds or thousands of samples per class[31,32]. In addition, it is difficult to obtain such a large number of samples in public databases[33], which is why traditional machine learning algorithms are more commonly used in transcriptomics sequencing data analysis than deep learning algorithms, as traditional machine learning algorithms do not require high sample size[34]. With the development of deep learning, data augmentation has been widely used in fields such as image and speech to generate additional training data and reduce overfitting and improve the generalization ability of the model[35]. In recent years, researchers have tried to use some generative models, such as GANs, to enhance transcriptomics data to improve classification performance[36,37]. WGAN is an improved GAN that uses the Wasserstein distance instead of the Jensen-Shannon divergence to measure between the generated distribution and the true distribution[38], which makes WGAN training more stable. WGAN-GP is an improved version of WGAN that adds gradient penalty to satisfy the Lipschitz constraint, further improving model performance. Compared with the original GAN, WGAN and WGAN-GP have significant improvements in generating sample quality and training stability[38]. Recently, researchers have used WGAN-GP to enhance transcriptomics data, and the results show that WGAN-GP-generated artificial samples can effectively improve the performance of subsequent classification models. Therefore, it is feasible and practical to enhance small-scale transcriptomics sequencing data with WGAN-GP, and then use Transformer models to classify biological processes and mine key genes.

In view of this, this study proposes a new model, named TransGeneSelector, which utilizes a Transformer model to classify biological processes and identify key genes. The model consists of a sample generation network based on WGAN-GP, a sample filtering network, and a classification network based on Transformer. The model first generates transcriptomic samples using the WGAN-GP sample generation network, and then filters out low-quality samples using the sample filtering network. Finally, the generated samples and original real samples are treated as sequence data input to the Transformer, which captures the global relationships between genes and performs the classification of biological processes. Subsequently, the importance of each gene is evaluated using SHAP (SHapley Additive exPlanations), an approach in explainable machine learning that aims to provide interpretable explanations for individual predictions made by any machine learning model[39,40].

The study found that TransGeneSelector can predict the seed state (dry seeds/or germinating seeds) with performance comparable to that of a Random Forest and identify genes at a more upstream regulatory level than those Random Forests mined, and the top 11 genes uniquely mined by TransGeneSelector were found to be related to the KAI2 signaling pathway, which is of great regulatory importance for germination-related genes. Therefore, our method TransGeneSelector has the following advantages: (1) It can analyze small sample size transcriptomic data and classify specific biological processes into two categories, and identify key genes involved in specific biological processes; (2) The Transformer model can capture important regulatory relationships between genes and identify upstream key genes that regulate specific biological processes. This study provides a practical tool for life science researchers to mine key genes from transcriptome data in specific biological processes.

## Result

**Overview of TransGeneSelector**

Overall, TransGeneSelector is a deep learning method for classifying biological processes and identifying key genes that combines multiple neural network models (Fig. 1). The method includes two larger networks and a smaller network. The larger networks include a sample generation model based on WGAN-GP and a classification model based on the Transformer architecture. The smaller network is an additional classifier network based on a fully connected neural network. The model works as follows: (1) fake sample generation (Fig.1a). First, the training set (cross-validation training set) is used to train the WGAN-GP model. After sufficient training, the WGAN-GP can generate fake transcriptomic gene expression samples (fake samples I).

Then, fake samples I and the real samples of the training set are input to the additional classifier for training, which improves the ability of the additional classifier to distinguish between fake and real samples. After training the additional classifier, we input the fake samples II generated by WGAN-GP to the additional classifier to filter out low-quality samples (samples with model output results lower than 0.1), and finally obtain high-quality samples, i.e., final fake samples. (2) Transformer classification (Fig.1b). We mix the final fake samples with the real samples of the training set and use them to train our Transformer classification model. To adapt to small-sample classification tasks, we simplify the Transformer model as much as possible, only keeping the encoder part. First, each sample's high-dimensional gene expression level is reduced to 72 dimensions by a simple full connected (FC) layer, instead of using a convolutional approach, to capture as much global expression information as possible in the simplest way. Then, the 72-dimensional vector is input to a stacked Attention head of 8 layers in the Transformer encoder after Positional Encoding. We take the first token's vector of the output and use it to evaluate the classification performance of the test set real samples. (3) SHAP method for key gene mining. We use the SHAP (SHapley Additive exPlanations) method to evaluate the impact of each gene on the trained Transformer classification model. Judging from Shapley values, we obtain genes that have the greatest impact on the classification result, which are considered key genes in a specific biological process.

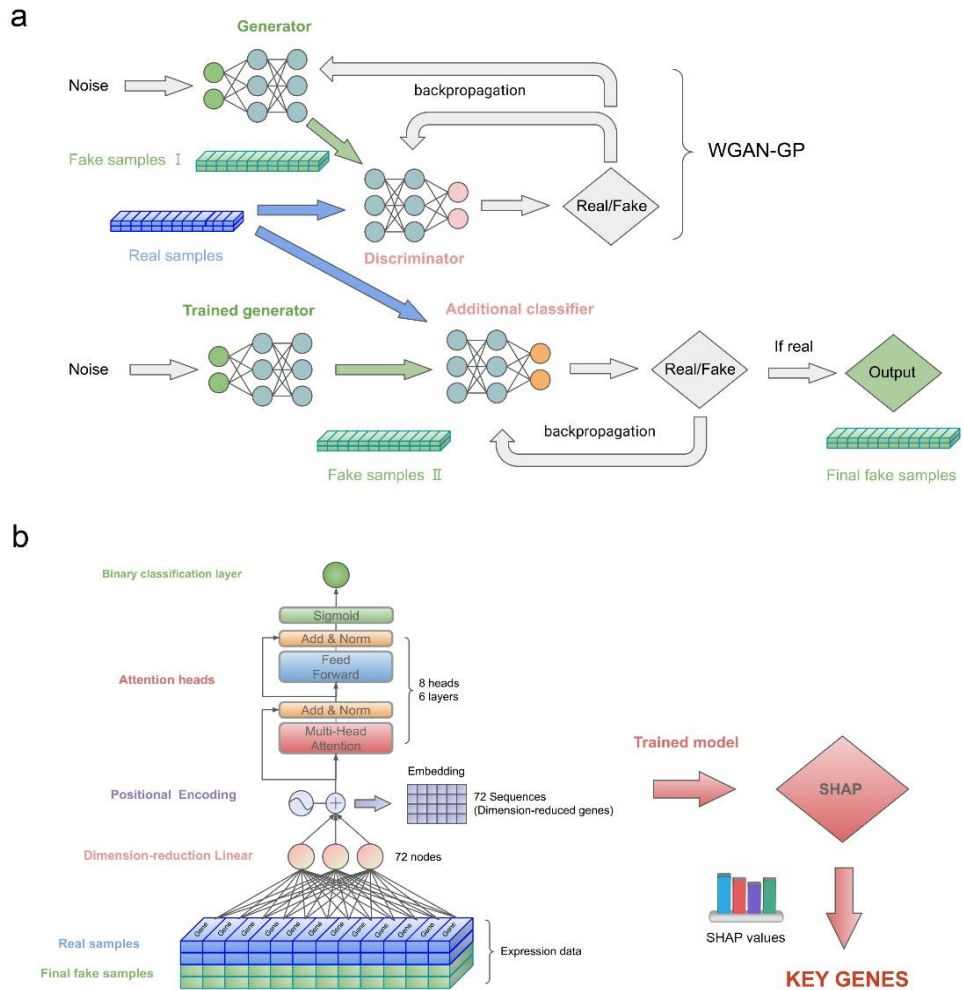

**Fig. 1** The overall frame and workflow of TransGeneSelector.
**a** Network for generating fake samples with the WGAN-GP and additional classifier sample filtering networks. The WGAN-GP model is trained with real sample data to generate fake gene expression samples, fake samples I, which are then combined with the training set of real samples and fed into the additional classifier for training to improve its ability to distinguish between fake and real samples. After the additional classifier is trained, we feed the fake samples II generated by WGAN-GP into the additional classifier to filter out low-quality samples, and finally obtain high-quality samples, namely final fake samples. **b** Transformer classification model and SHAP method for mining key genes. We mix the final fake samples with the training set of real samples and use them to train our Transformer classification model, which we have simplified as much as possible to suit small-sample classification tasks. We applied the Sigmoid function to the first token of the output. Finally, we use the SHAP (SHapley Additive exPlanations) method to evaluate the impact of each gene on the trained Transformer classification model, and obtain the genes that have the greatest impact on the classification results, achieving the mining of key genes for a specific biological process.

**TransGeneSelector achieves high classification performance in small samples**

First, we trained the TransGeneSelector using a gene expression dataset of *Arabidopsis thaliana* dry seeds (negative samples, 36 samples) and germinating seeds (positive samples, 43 samples), which consisted of 79 samples. The training of the WGAN-GP module was stable, with good performance after 3800 training epochs, with the Generator and Discriminator's Loss showing a convergence state (Fig. 2a). Moreover, after training for 3800 epochs with these parameters, the Fréchet Inception Score value reached the lowest level (Fig. 2b), and the distribution of generated samples was close to the distribution of real samples (Fig. 2c). Therefore, we performed subsequent network training based on these parameters.

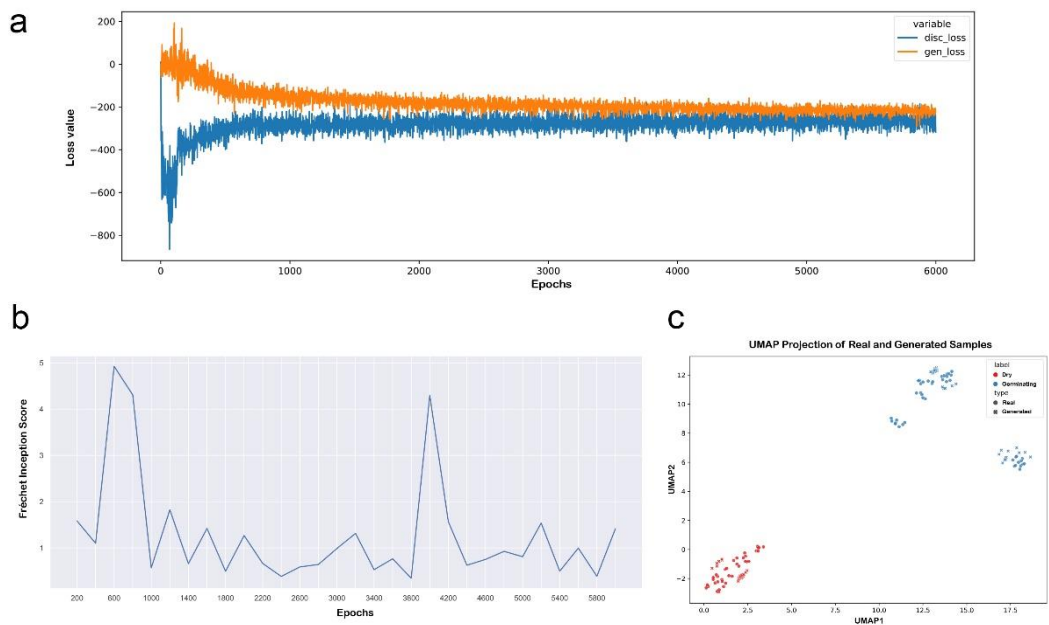

**Fig. 2** Training process and sample quality evaluation of WGAN-GP.
**a** Relationship between the number of training epochs and the Loss value of WGAN-GP. The disc_loss is the loss of the discriminator, and the gen_loss is the loss of the generator. **b** Relationship between the number of training epochs and the Fréchet Inception Score of WGAN-GP. **c** UMAP visualization of the generated samples and the real samples. Different colors represent seeds in different states (dry seeds or germinating seeds), and different shapes represent the generated samples and the real samples.

After thoroughly training the WGAN-GP, we utilized the real samples from the training set and the same number of fake samples generated by WGAN-GP to train an additional classifier network for 50 epochs. This allowed the network to fully learn the distinct characteristics of genuine and fake samples. Following the completion of the additional classifier network's training, we set the threshold to 0.1 and selected 0, 100, 400, 700, 1000, 1300, 1600, 1900, 2200, and 2500 high-quality fake samples generated by WGAN-GP. Then, we combined the fake samples and real samples (cross-validation training set) and fed them into the Transformer network of TransGeneSelector for training of the Transformer classification model. Following the training process, we applied the validation set to evaluate the accuracy, precision, recall, and F1 index. We performed 5-fold cross-validation for each sample amount. The results (Fig. 3a) indicate that, with the fake samples amount getting greater, regardless of whether an additional classifier is used, all four indicators of cross-validation exhibit a trend of increasing and decreasing later, with the accuracy, precision, and F1 index of the model that includes an additional classifier and 100 fake samples achieving the highest values of 0.974, 1.000, and 0.976, respectively (Fig. 3a). This suggests that the WGAN-GP network enhances the performance of the Transformer classification model, and the optimal number of fake samples in this case is 100, which is close to the amount of real samples, namely 63. Furthermore, it is evident that the model using an additional classifier outperforms the model without an additional classifier in most cases, and the standard deviation of most indicators is reduced, particularly in the case of 100 fake samples, when the standard deviation is the lowest (Fig. 3a), indicating that the additional classifier network stabilizes and enhances the reliability of the model training process.

We also used Random Forests to classify germinating seeds and dry seeds using the same dataset. We used the Wrapper method to select 200 parameters (number of genes) of n_features_to_select uniformly spaced between 1 and 500 for feature engineering and classification evaluation in Random Forests, and used cross-validation to test the accuracy of the Random Forest model. The results (Fig. 3b) showed that the accuracy of Random Forests could reach 1.000 under certain feature (genes) numbers, indicating that traditional machine learning algorithms do have advantages in classification on small sample datasets. However, considering that TransGeneSelector is a deep learning-based model, its highest accuracy, precision, and F1 index reached 0.974, 1.000, and 0.976, respectively, which were very close to that of the Random Forests model. Therefore, our proposed method can exhibit performance comparable to that of Random Forests on small sample datasets.

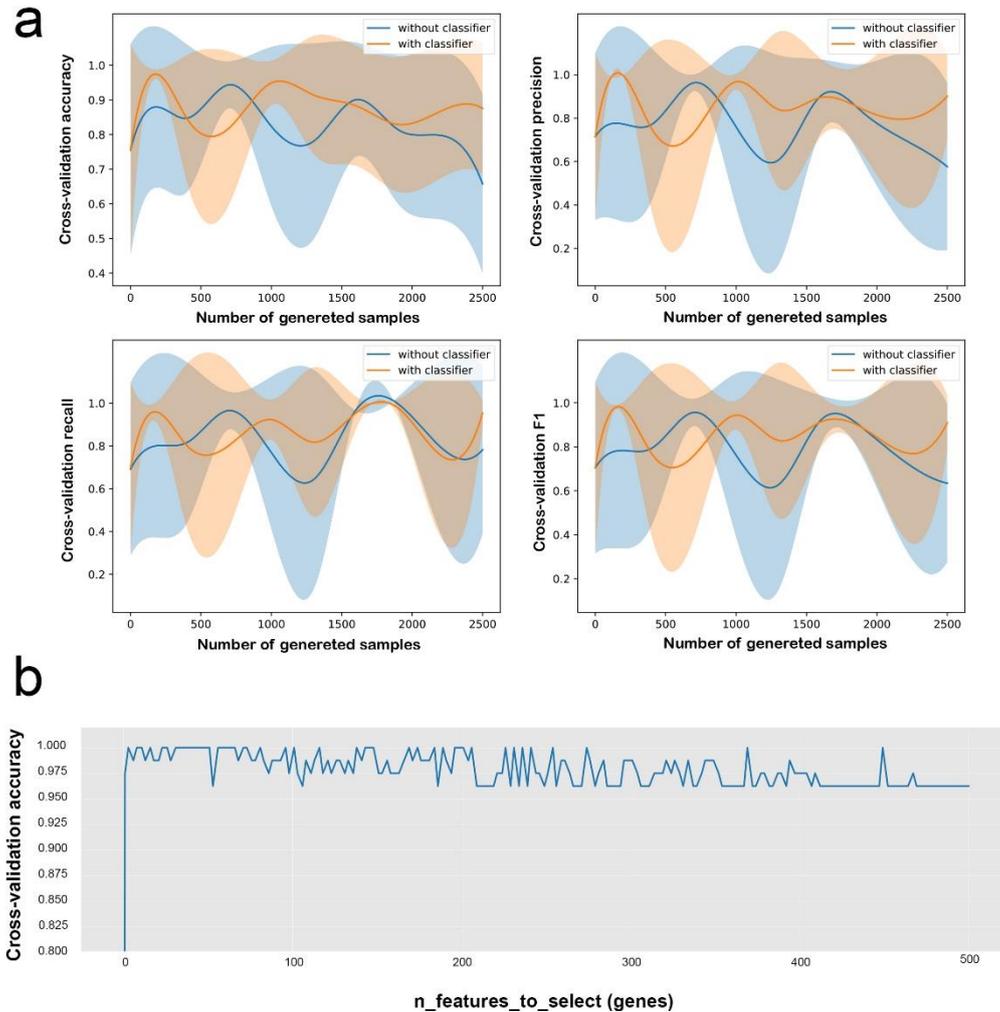

**Fig. 3** Comparison of classification performance of TransGeneSelector and Random Forests.
**a** Classification performance of TransGeneSelector under different numbers of generated samples. The blue color represents the cross-validation performance of the model without additional classifier filtering, and the orange represents the cross-validation performance of the model with additional classifier filtering. The shaded areas represent the error bands. **b** Performance of Random Forests classifier with different numbers of genes selected by wrapper method. n_features_to_select is a parameter of RandomForestClassifier module, which is set uniformly spaced between 1 and 500 with a value of 200.

**TransGeneSelector has powerful key gene mining capability**

We evaluated the feature importance of TransGeneSelector's Transformer network using the SHAP method, reflecting the impact of each gene on the model's output, and mining key genes involved in seed germination. We also used the Wrapper method in Random Forests to screen key genes involved in seed germination. We compared the expression pattern of the screened genes in all samples (Fig. 4a). For Random Forests, we chose n_features_to_select parameters (gene numbers) with cross-validation classification accuracies of 1.0, they were 11, 51, 148, and 449. and then compared the same number of genes selected by SHAP for TransGeneSelector.

The comparison of expression patterns (Fig. 4a) showed that, first, when the number of genes targeted for selection was low (11 and 51), the genes randomly selected by Random Forest were significantly different in expression patterns between dry seeds and germinating seeds, and the expression levels within a single group (dry seeds or germinating seeds) were very consistent, which can be clearly distinguished from the other group (Fig. 4a). This consistency was also significantly higher than that of genes selected by TransGeneSelector with the same gene numbers. However, as the number of genes targeted for selection increased to 148, the expression patterns of genes selected by Random Forest became chaotic, and the difference between the expression patterns of germinating seeds and dry seeds became less obvious (Fig. 4a). This phenomenon became more prominent as the number of genes targeted for selection increased to 449 (Fig. 4a). On the other hand, TransGeneSelector showed the opposite trend. Although the expression patterns of the top 11 genes and 51 genes selected by TransGeneSelector were not as consistent within each group (dry seeds or germinating seeds) as those of genes selected by Random Forest, they became increasingly consistent as the number of genes targeted for selection increases, and the difference in expression patterns between germinating seeds and dry seeds became much more obvious (Fig. 4a). This phenomenon was also true when the number of genes targeted for selection reached 449. This indicates that TransGeneSelector and Random Forest are different in the characteristics for key gene mining. If the expression patterns are the only index used to evaluate the effectiveness of key gene selection,

TransGeneSelector is more advantageous than Random Forest in selecting genes with a large number of genes.

We used the MERLIN gene regulatory network construction algorithm to explore the regulatory relationships between all 885 genes mined by Random Forest and TransGeneSelector (Fig. 4b). The results showed that when using larger datasets that were not from seed and not related to seed germination to infer the network, the number of regulatory relationships between genes were higher, and the types of regulatory relationships across the two methods (from genes of Random Forest to those of TransGeneSelector, or from genes of TransGeneSelector to those of Random Forest) were evenly distributed (50%)(Fig. 4b). However, when using the dry seed and seed germination datasets of our study to infer the network, the total number of regulatory relationships between genes decreased, but the proportion of regulatory relationships across the two methods that went from genes of TransGeneSelector to those of Random Forest (66.7%) was significantly higher than the proportion that went from genes of Random Forest to those of TransGeneSelector (33.3%)(Fig. 4b). When the dataset used is related to germination, the gene regulatory network using the dataset can better capture the regulatory relationships related to seed germination, and as a result in this case, the majority of the regulatory relationships were from genes of TransGeneSelector to those of Random Forest, we believe that TransGeneSelector can identify genes with greater upstream regulatory effects on seed germination than Random Forest.

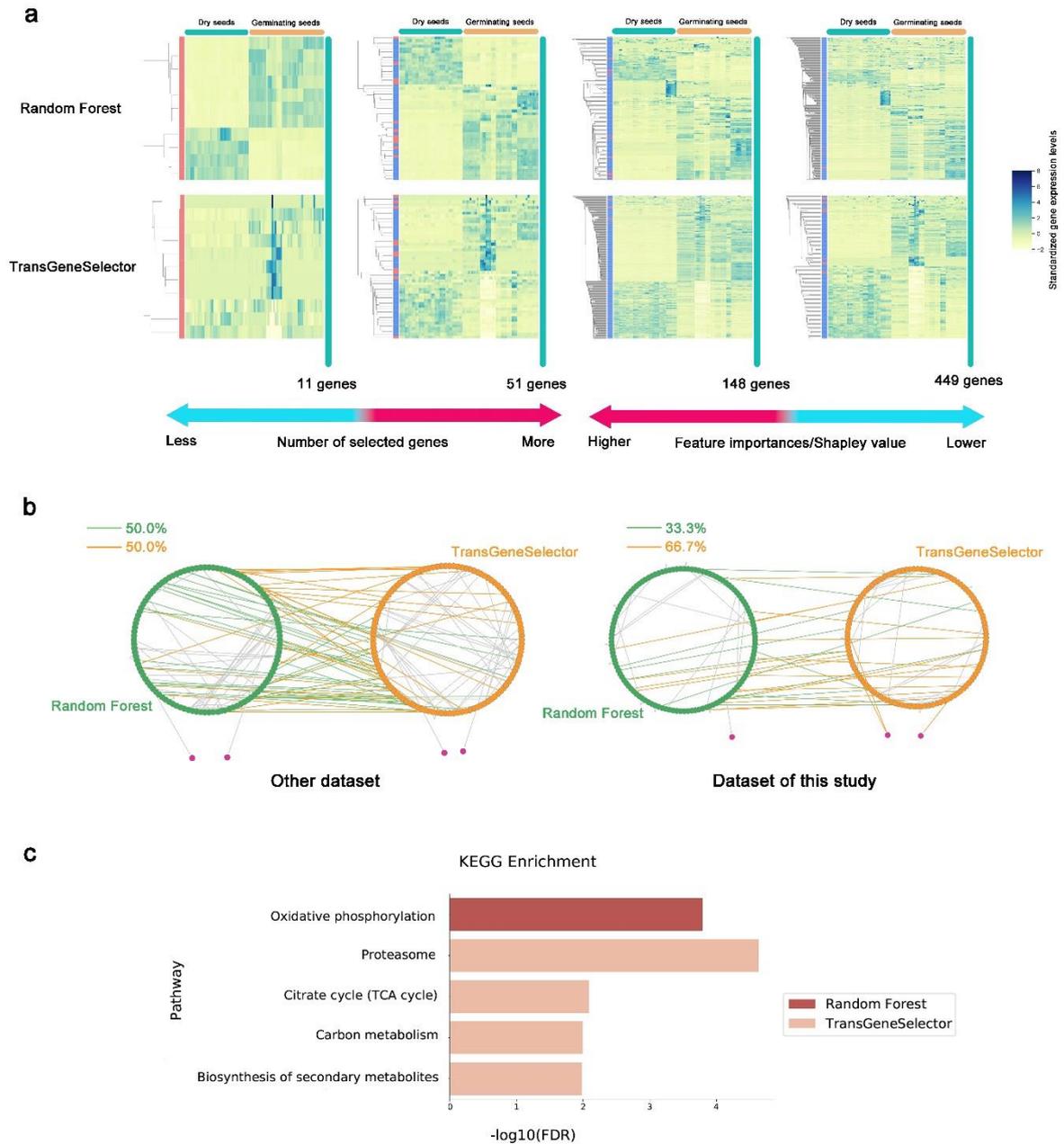

**Fig. 4** Comparison of TransGeneSelector and Random Forests in gene mining.
**a** Heatmap of transcriptomics expression profiles of genes selected by TransGeneSelector and Random Forests with different numbers of genes. The darker the color, the higher the expression level. TransGeneSelector selects genes based on Shapley values in descending order, while Random Forests selects genes based on setting different numbers of n_features_to_select. **b** Gene regulatory relationship between genes selected by TransGeneSelector and Random Forests. Green nodes represent genes selected by Random Forests, orange nodes represent genes selected by TransGeneSelector, and purple nodes represent genes selected by both methods. Green lines represent the regulatory relationships from genes selected by Random Forests to those selected by TransGeneSelector, and orange lines represent the regulatory relationships from genes selected by TransGeneSelector to those selected by Random Forests. **c** Pathway enrichment analysis of genes selected by TransGeneSelector and Random Forests in KEGG.

Kyoto Encyclopedia of Genes and Genomes (KEGG) focuses on the relationships between genes and metabolic pathways, making it suitable for capturing the interactions and functions of genes. We further annotated and enriched each 449 genes mined by Random Forests and TransGeneSelector (Fig. 4c), and the KEGG enrichment results showed that the genes extracted by Random Forests were significantly enriched only in one pathway of Oxidative phosphorylation, while the genes extracted by TransGeneSelector were significantly enriched in four pathways, including Proteasome, Citrate cycle (TCA cycle), Carbon metabolism, and Biosynthesis of secondary metabolites (Fig. 4c). According to the KEGG enrichment results, the genes mined by the two methods had both similarities and differences. Firstly, both enriched pathways included pathways related to energy metabolism, such as Oxidative phosphorylation, Citrate cycle (TCA cycle), and Carbon metabolism. However, the genes extracted by Random Forests were enriched in the Oxidative phosphorylation pathway, which is mainly involved in energy metabolism and ATP synthesis. This pathway is downstream in the energy generation process, as it uses the products of other metabolic pathways, such as NADH and FADH2, to generate ATP[41]. On the other hand, the genes extracted by TransGeneSelector were enriched in pathways like the Citrate cycle (TCA cycle) and Carbon metabolism. These pathways are involved in central carbon metabolism processes, such as glycolysis and the pentose phosphate pathway, which generate precursor metabolites and reducing equivalents (NADH and FADH2) that are used in the Oxidative phosphorylation pathway[42]. Therefore, these pathways can be considered upstream of the Oxidative phosphorylation pathway. In addition, the genes significantly enriched in the Proteasome pathway are of key values for seed germination, as seed germination often involves the degradation of stored proteins in the proteasome[43-45], and the Biosynthesis of secondary metabolites involves the synthesis of various secondary metabolites, which play essential roles in various biological processes, which suggests the beginning of life diversity and is closely related to seed germination[46-48]. In this study, the genes of these important pathways were all mined by TransGeneSelector. Therefore, in summary, TransGeneSelector can identify a more upstream and representative set of genes related to seed germination than Random Forests.

**TransGeneSelector can identify upstream key regulatory genes**

Both the top 11 genes and the top 51 genes with the largest Shapley values mined by TransGeneSelector were not as consistent within the germination group or dry seed group as the corresponding number of genes mined by the Random Forests. Is this a weakness of TransGeneSelector, or is it actually its unique feature, and do these genes have greater significance? To answer this question, we first removed the 11 genes selected by TransGeneSelector from our original dataset and trained the TransGeneSelector model using the remaining dataset with the same parameters. We then observed the model's classification performance (Fig. 5a) and found that when we removed these 11 genes, the model's performance at the position (100 samples generated by WGAN-GP) with the strongest performance trained using the original full dataset significantly decreased (Fig. 5a). The accuracy, precision, recall, and F1 score decreased from 0.974, 1.000, 0.955, and 0.976 to 0.887, 0.800, 0.737, and 0.767, respectively. Additionally, when the 11 genes were removed the standard deviation of the four indicators was significantly larger than before (Fig. 5a). Therefore, this result indicates that the 11 genes selected by TransGeneSelector have important implications for the model's classification performance, suggesting that these genes may play important roles.

To explore the specific functions of these 11 genes, we analyzed their expression patterns in all samples of the dataset (Fig. 5b), and the results showed that most of these 11 genes were highly expressed only in dark germinating *Arabidopsis thaliana* seeds, including AT4G03050, AT4G33320, AT2G12260, AT1G58684. Three other genes, AT3G56820, AT5G16560, and AT4G17060, showed low expression characteristics also only in dark germinating *Arabidopsis thaliana* seeds. In other cases, such as dry seeds, ABA-treated germinating seeds, NaCl-treated germinating seeds, and normal germinating seeds, these genes had little expression variation (Fig. 5b). More importantly, the expression patterns of these genes were completely consistent with the expression patterns of genes *KAI2* and *SMAX1*, which play important role in KAI2 pathway that regulates seed germination under dark condition, indicating that these genes may all be involved in *Arabidopsis thaliana* seed germination in the dark.

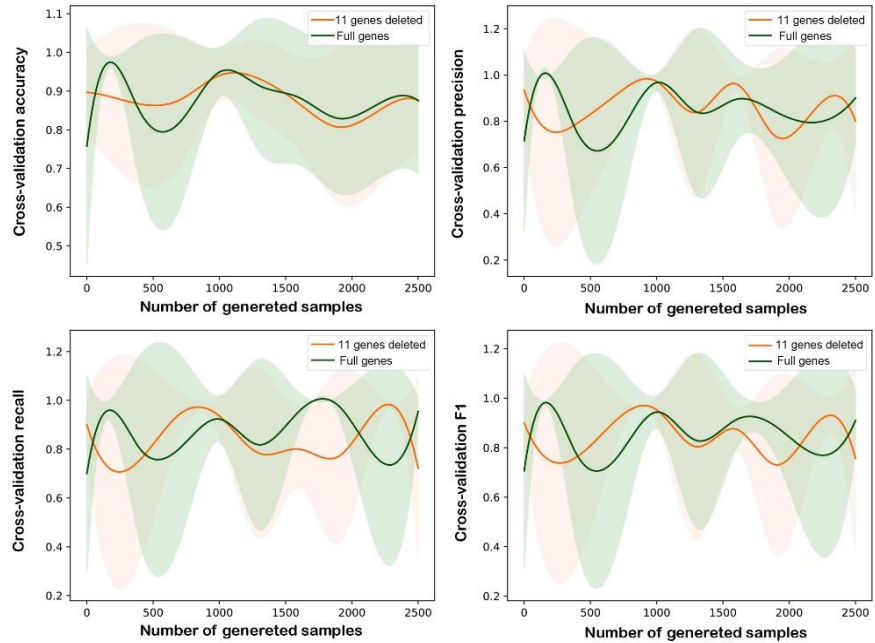

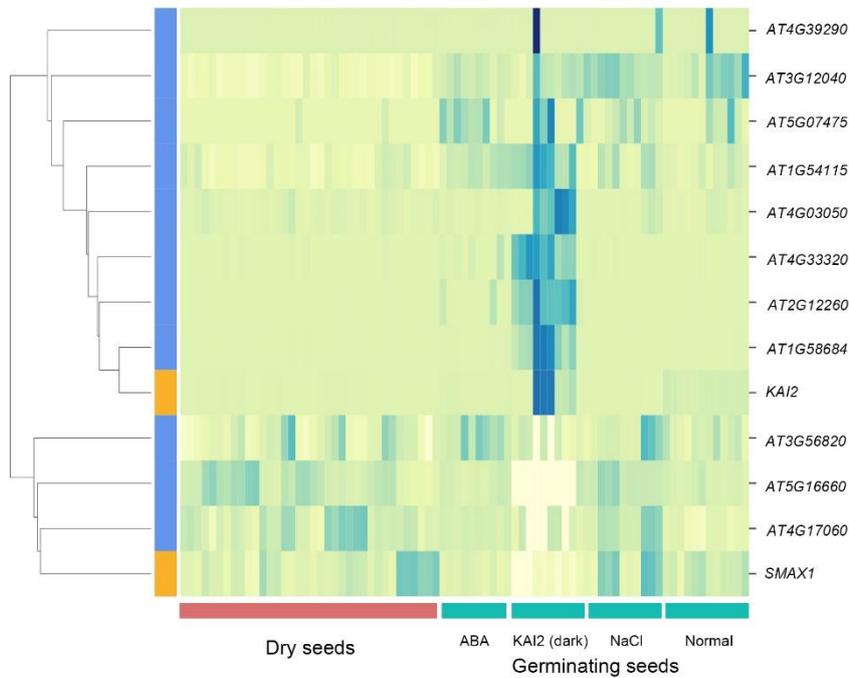

**Fig. 5** Impact of the top 11 genes selected by TransGeneSelector on model performance and their transcriptomics expression profiles.

**a** Impact of the top 11 genes selected by TransGeneSelector on model performance. Orange represents the cross-validation results of TransGeneSelector trained on the data set without the top 11 genes, and green represents the cross-validation results of TransGeneSelector trained on the original data set. **b** Transcriptomics expression profiles of the top 11 genes selected by TransGeneSelector, where the darker the color, the higher the expression level. The blue bars represent the genes selected by TransGeneSelector, and the orange bars represent the genes in the KAI2 pathway. Dry seeds represent the samples of dry seeds of *Arabidopsis thaliana*, ABA represents the germinating seeds of *Arabidopsis thaliana* related to ABA treatment, KAI2 (dark) represents the germinating seeds of *Arabidopsis thaliana* in the dark with KAI2 pathway-related genes highly activated, NaCl represents the germinating seeds of *Arabidopsis thaliana* under salt stress, and Normal represents the germinating seeds of *Arabidopsis thaliana* under normal conditions.

To determine whether these genes are involved in *Arabidopsis thaliana* seed germination in the dark, we germinated seeds of *Arabidopsis thaliana* Col-0 under different light intensities of 0h (dry seeds), 12h, 24h, and 28h, and then measured the expression levels of the 10 genes among the 11 genes by RT-qPCR. The results (Fig. 6) showed that the expression patterns of these genes in seeds germinating under different light conditions were surprisingly consistent, with high expression in the seeds that germinated in the dark for 24 and 48 hours, and under light for 48 hours, and low expression in other cases (Fig. 6). Moreover, the expression patterns of these 10 genes were consistent with the expression patterns of *KAI2* gene in the dark and completely consistent with the expression patterns of *SMAX1* gene in all cases. This suggests that the genes selected by TransGeneSelector with the largest Shapley values in this study are indeed involved in seed germination in the dark and are related to the KAI2 pathway. Studies have shown that activating the KAI2 signaling pathway for dark germination of *Arabidopsis thaliana* seeds can increase the levels of gibberellin (GA) and ethylene signals and suppress abscisic acid (ABA) signals. This hormone combination state is beneficial for seed germination, and the KAI2 signaling pathway can replace light to activate GA and ethylene signals, suppress ABA signals, similar to the hormone mechanism of seed germination under light, thus promoting seed germination in the dark[49,50]. Therefore, gene of KAI2 signaling pathway has important regulatory effects on genes and pathways related to germination, and the 11 genes with the largest Shapley values mined by TransGeneSelector in this study are exactly related to the KAI2 signaling pathway, so these genes all have important regulatory effects on seed germination and have a significant regulatory impact on other genes related to seed germination. In this study, Random Forests cannot mine these important genes (Fig. 4a), we speculate that these genes with significant effects can only be identified by the characteristics of the Transformer network of TransGeneSelector, which captures the mutual regulatory relationships between genes and ultimately identifies these genes that have important regulatory effects on seed germination and other genes related to seed germination. Therefore, we demonstrate that TransGeneSelector is a deep learning method that can identify upstream key regulatory genes in biological processes.

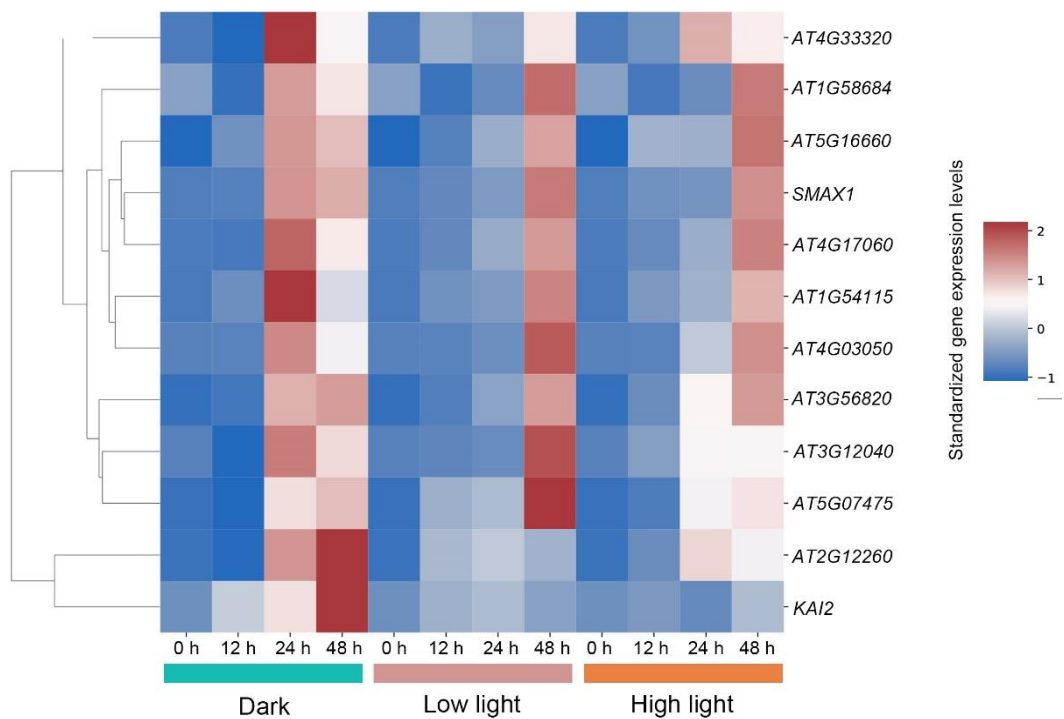

**Fig. 6** RT-qPCR quantitative results of the top 11 genes selected by TransGeneSelector in *Arabidopsis thaliana* seeds under different germination conditions.
SMAX1 and KAI2 are two genes in the KAI2 pathway, and the other genes are the top 11 genes selected by TransGeneSelector. Dark refers to seed germination under full-black conditions, i.e., aluminum foil wrapped full-black germination conditions, Low light refers to seed germination under low light conditions of 100 μmol photons photosynthetic light, and High light refers to seed germination under high light conditions of 200 μmol photons photosynthetic light.

# Dicussion

Gene mining plays an important role in the field of life sciences, but existing traditional machine learning methods cannot consider the mutual regulatory relationships between genes, while deep learning methods cannot perform well in small sample size scenarios. Therefore, this study proposes TransGeneSelector, a deep learning method that can mine upstream key regulatory genes for certain biological process with small sample size dataset, which makes up for the deficiencies of traditional machine learning and deep learning methods in excavating key genes from small samples, and has been successfully applied to the excavation of key upstream regulatory genes for seed germination of *Arabidopsis thaliana.*

First, by using data augmentation with the WGAN-GP network, data augmentation quality control with the additional classify network, and a simplified designed Transformer network structure, we successfully achieved high-performance classification of dry seeds and germinating seeds in a small data set of 79 samples using TransGeneSelector. The classification performance of TransGeneSelector can directly compete with that of Random Forests, successfully realizing the application of deep learning to classification task of small sized dataset.

Then, by using the multi-head self-attention mechanism of the Transformer network to globally capture the expression status of each gene, and the SHAP interpretable machine learning method to evaluate the importance of each gene, TransGeneSelector successfully mined a certain number of genes. Through the construction of gene regulatory networks and enrichment analysis with KEGG, we found that the genes mined by TransGeneSelector are with more upstream regulatory ability and can better reflect the physiological process of seed germination than those identified by the Random Forests method.

More importantly, it is worth noting that TransGeneSelector mined 11 genes with the highest Shapley values that exhibited special expression patterns in all samples, and qPCR experiments confirmed that their expression patterns were consistent with the two key genes *KAI2* and *SMAX1* of the KAI2 pathway in seeds under all kinds of light condition for seed germination, especially under dark condition. *KAI2* and *SMAX1* play important role in KAI2 pathway that regulates seed germination under dark condition[49-52], This confirmed that the 11 genes were all located in the KAI2 pathway, a pathway of strong regulatory ability for seed germination and had important regulatory effects on seed germination and seed germination-related genes through its influence on ABA and GA pathway for seed germination [49-52]. Thus, these most important 11 genes mined by TransGeneSelector have a significant regulatory impact on other genes related to seed germination, and could not be mined by Random Forests method. In addition, the 11 genes we identified have never been reported to be associated with seed germination, and their roles in germination deserve further investigation. This study is also the first to discover that the Transformer network structure can identify such a special gene group in transcriptomic data, indicating that gene mining based on the Transformer network structure is valuable and deserves further in-depth study. In summary, we believe that TransGeneSelector, which has a WGAN-GP structure and a Transformer network structure, is an excellent practical tool for key gene mining in small sample transcriptomic data.

Although TransGeneSelector performed extremely well in terms of classification performance and key gene mining in small sample data, there is still room for improvement. First, TransGeneSelector is a complex system involving three separate neural networks, so it requires more complex human operations when training each part separately. For example, in TransGeneSelector, the WGAN-GP module needs to determine the best sample-generating parameters in advance, and the additional classifier needs to choose suitable thresholds to make the Transformer module perform best. This complexity requires further optimization in the future. In addition, due to the application for small sized data, TransGeneSelector was designed to perform binary classification tasks only, so further research is needed in multi-classification tasks.

In summary, we propose and evaluate the performance of TransGeneSelector in key gene mining in small sample transcriptomic data, which is a tool with strong practicality for life science researchers to mine key genes involved in certain biological processes from transcriptomic data. It can mine upstream key regulatory genes involved in life processes, and our study provides a new perspective for the research of gene mining using Transformer.

## Materials and Methods

**Data description and processing**

Our data were collected from the NCBI GEO (Gene Expression Omnibus) database (https://www.ncbi.nlm.nih.gov/gds/) and Expression Atlas database (https://www.ebi.ac.uk/gxa/experiments). For training and testing of TransGeneSelector and Random Forest, experiments are GSE116069, GSE161704, GSE163057, GSE167244, and GSE179008 from NCBI, including raw counts of 79 samples from dry seeds and germinating seeds, with 43 positive samples and 36 negative samples. We calculated the TPM of each gene based on its length and raw counts, and retained genes that were included in all samples. We set the samples from the germinating group as positive samples (1) and the samples from the non-germinating group as negative samples (0). For network analysis, experiments are E-CURD-1, E-GEOD-30720, E-GEOD-52806, E-GEOD-64740, E-MTAB-4202, E-MTAB-7933, and E-MTAB-7978 from Expression Atlas and GSE199116 from NCBI, which included 268 samples unrelated to seed germination. We calculated the TPM of each gene, and retained genes that were included in all samples.

**TransGeneSelector framework**

TransGeneSelector includes three neural networks, respectively, a sample generation network based on Wasserstein GAN with Gradient Penalty (WGAN-GP), an additional classifier network with a fully connected neural network architecture, and a classification network based on the Transformer architecture.

WGAN-GP, the sample generation network of TransGeneSelector, is an improvement over the original Wasserstein GAN (WGAN) [38,53] that addresses the limitations of the original model by using a gradient penalty instead of weight clipping to enforce the Lipschitz constraint. This results in more stable training and better convergence

properties. For the original Wasserstein GAN, the loss function of the discriminator (critic) is defined as:

$$L_{Critic} = E_{x \sim p_r}[f_w(x)] - E_{z \sim p_z}[f_w(g_\theta(z))]$$

Where $f_w$ is the discriminator (critic), $g_\theta$ is the generator, $p_r$ is the real data distribution.

$p_z$ is the noise distribution.

For the generator, the loss function is defined as:

$$L_{Generator} = -E_{z \sim p_z}[f_w(g_\theta(z))]$$

The WGAN model is trained by solving the following optimization problem:

$$\min_\theta \max_w [L_{Critic}]$$

The WGAN-GP loss function is defined as:

$$V_{WGAN-GP}(f_w, g_\theta) = V_{WGAN}(f_w, g_\theta) + \lambda E_{\hat{x} \sim P_{\hat{x}}}[(\| \nabla_{\hat{x}} f_w(\hat{x}) \|_2 - 1)^2]$$

Where $V_{WGAN}(f_w, g_\theta)$ is the original WGAN loss function, which is designed to address the limitations of the standard GAN loss function by using the Wasserstein distance instead of the Jensen-Shannon divergence. It consists of two parts: one for the discriminator (critic) and one for the generator. $f_w$ is the discriminator (also called critic) in the WGAN-GP model. $g_\theta$ is the generator in the WGAN-GP model. $\lambda$ is a hyperparameter that controls the strength of the gradient penalty term. $\hat{x} \sim P_{\hat{x}}$ is the expectation over random samples $\hat{x}$ drawn from the distribution $P_{\hat{x}}$. In WGAN-GP, $\hat{x}$ is a randomly weighted average between a real data point and a generated data point. $(\| \nabla_{\hat{x}} f_w(\hat{x}) \|_2 - 1)^2$ is the gradient penalty term. It penalizes the squared difference between the gradient norm of the discriminator with respect to its input $\hat{x}$ and the target norm value 1. The purpose of this term is to enforce the Lipschitz constraint on the discriminator, which helps to stabilize the training and improve convergence properties.

The WGAN-GP model is trained by solving the following optimization problem:

$$\min_\theta \max_w [V_{WGAN-GP}(f_w, g_\theta)]$$

After generating the fake samples, the additional classifier network with a fully connected neural network architecture is used to filter out the fake samples and obtain high-quality samples. The network architecture consists of several fully connected layers (also known as linear layers) with Rectified Linear Unit (ReLU) activation functions in between, followed by a final linear layer with a Sigmoid activation function to output a probability value between 0 and 1. Here's the mathematical representation of the additional classifier network:

$$h_1 = \text{ReLU}(W_1 x + b_1)$$
$$h_2 = \text{ReLU}(W_2 h_1 + b_2)$$
$$h_3 = \text{ReLU}(W_3 h_2 + b_3)$$
$$h_4 = \text{ReLU}(W_4 h_3 + b_4)$$
$$h_5 = \text{ReLU}(W_5 h_4 + b_5)$$
$$y = \text{Sigmoid}(W_6 h_5 + b_6)$$

Where $h_i$ represents the output of the $i^{th}$ hidden layer, $W$ and $b$ are the weight matrix cnd bids vector for the $i^{th}$ layer, respectively. ReLU is the Rectified Linear Unit activation function, defined as:

$$\text{ReLU}(x) = \max(0, x)$$

Sigmoid is the Sigmoid activation function, given an input $x$, the output $\sigma(x)$ of the Sigmoid function is calculated as:

$$\sigma(x) = \frac{1}{1 + e^{-x}}$$

The network takes an input vector and passes it through the layers to produce a single output probability value. The output value can be thresholded to obtain the high-quality generated samples.

After the above sample-generating processes, the generated samples and real samples for seeds under both kinds of conditions (germinating or dry seeds) were altogether input into the Transformer network for biological process classification. It is start by using a fully connected network to reduce the dimensionality of the gene expression data for the numerous number of genes. The output of this step is a lower-dimensional representation of the input genes. Given an input expression value $x$, the output $y$ of a fully connected layer with weights $W$ and biases $b$ is calculated as

$$y = x^T W + b$$

The lower-dimensional representation is then positional encoded to provide the Transformer network with information about the order of representation. The formula used for calculating the positional encoding values is as follows:

$$PE_{(pos,2i)} = sin(pos/10000^{2i/d_{model}})$$
$$PE_{(pos,2i+1)} = cos(pos/10000^{2i/d_{model}})$$

where $pos$ is the position of the word in the sequence, $i$ is the index of the dimension pair, and $d_{model}$ is the dimension of the input embeddings.

When the lower-dimensional representation of the gene expression data is positional encoded, it is then fed into the Transformer Encoder. The Encoder processes the input sequence and produces a continuous representation, or embedding, of the input. The Transformer Encoder consists of multiple self-attention and feed-forward layers, allowing the model to process and understand the input sequence effectively. The multi-head self-attention mechanism in the encoder allows the model to attend to different parts of the input sequence simultaneously. It computes multiple attention outputs in parallel and then concatenates them before passing them through a linear transformation. The multi-head attention can be represented as:

$$\text{MultiHead}(\mathbf{Q}, \mathbf{K}, \mathbf{V}) = [\text{head}_1, \ldots, \text{head}_h]\mathbf{W}_0$$

$$\text{head}_i = \text{Attention}(\mathbf{Q}\mathbf{W}_i^Q, \mathbf{K}\mathbf{W}_i^K, \mathbf{V}\mathbf{W}_i^V)$$

Here, $\mathbf{Q}, \mathbf{K}, \mathbf{V}$ represent the query, key, and value matrices, respectively, and $\mathbf{W}$ are the learnable parameter matrices. The scaled dot-product attention computes the attention scores by taking the dot product of the query and key matrices, dividing the result by the square root of the key vector dimension, and then applying a softmax function:

$$\text{Attention}(\mathbf{Q}, \mathbf{K}, \mathbf{V}) = \text{softmax}(\frac{\mathbf{Q}\mathbf{K}^T}{\sqrt{d_k}})\mathbf{V}$$

After the multi-head self-attention mechanism, the output is passed through a position-wise feed-forward network, which consists of two linear layers with a ReLU activation

function in between. The position-wise feed-forward network can be represented as follows:

$$\text{FFN}(x) = \text{ReLU}(x^T W_1 + b_1)W_2 + b_2$$

Where $x$ is the input, $W_1$ and $W_2$ are the weight matrices, and $b_1$ and $b_2$ are the bias terms.

Finally, residual connections and layer normalization are applied after both the multi-head self-attention and position-wise feed-forward network to stabilize the training process and improve the model's performance. Residual connections are used to allow gradients to flow through a network directly. The residual connection formula is:

$$\text{Residual}(x) = F(x) + x$$

where $F(x)$ is the output of the previous layer and $x$ is the input

Layer normalization is applied to stabilize the training process. The layer normalization formula is:

$$\text{LayerNorm}(x) = \frac{x - \mu^l}{\sigma^l} * \gamma + \beta$$

where $\mu^l$ and $\sigma^l$ are the mean and standard deviation of the layer, respectively, and $\gamma$ and $\beta$ are learnable scale and shift parameters.

After processing the positional encoded lower-dimensional representation of the gene expression data through the Transformer Encoder, we use the first token of the output for classification, which is considered to contain the most relevant information for classification. We applied the Sigmoid function to the first token of the Encoder output, which is defined as :

$$\sigma(x) = \frac{1}{1 + e^{-x}}$$

The Binary Cross-Entropy (BCE) loss function is then used as loss function for binary classification of dry seeds and germinating seed. It measures the dissimilarity between the predicted probability distribution and the true binary labels of a dataset. The BCE loss function is particularly useful when the output is a probability value between 0 and 1.

$$BCE = -\frac{1}{N}\sum_{i=1}^{N}[y_i\log(\hat{y}_i) + (1-y_i)\log(1-\hat{y}_i)]$$

Where $N$ is the number of samples in the dataset. $y_i$ is the true binary label of the $i^{th}$ sample (1 for the positive class and 0 for the negative class). $\hat{y}_i$ is the predicted probability of the $i^{th}$ sample belonging to the positive class.

To take into account the non-linearity of the activation functions and maintain the standard deviation of the activations around 1, we used the Kaiming Uniform Initialization [54] for initializing the weights of the Transformer network, the Kaiming Uniform Initialization initializes the weights from a uniform distribution $U(-a, a)$, where:

$$a = \sqrt{\frac{6}{n_l}}$$

**Benchmarking quantification and statistics analysis**

To determine the default parameters of TransGeneSelector in the neural networks, we performed a grid search test for the parameters of each TransGeneSelector part neural network using the dry seed and germinating seed datasets. Here, WGAN-GP set the epochs from 200 to 6000 in steps of 200, with a combination of learning rate (0.1, 0.01, 0.001), the loss curve, the Fréchet Inception Distance (FID), and Uniform Manifold Approximation and Projection (UMAP) visualization were used to evaluate the best parameters of the WGAN-GP model. The additional classifier set the epochs (100, 150) and the learning rate (0.1, 0.01) combinations, respectively. The Transformer network set the embedding and header number combinations (72/8, 240/8, 72/16, and 240/16), the learning rate (0.1, 0.01, 0.001), and the training period (7, 21, and 35) combinations. The best parameter combination was determined by the validation set loss values.

To compare the performance of TransGeneSelector with Random Forest, we trained a Random Forest using the same dry seed and germinating seed datasets, and used grid search method to combine the parameter n_estimators (10, 100) and 200 parameters of n_features_to_select uniformly spaced between 1 and 500. The best parameter combination was determined based on the accuracy of the model.

To evaluate and compare the performance of trained TransGeneSelector and Random Forest, we used the 5-fold cross-validation method to assess the accuracy, precision, recall rate, and F1 score of TransGeneSelector, and we used the 5-fold cross-validation method to assess the accuracy of Random Forest, since it had a cross-validation accuracy of 100% in many occasions, other metrics were not calculated.

We applied SHAP (SHapley Additive exPlanations) [39] to mine important genes through Transformer network of TransGeneSelector by calculating the contribution of each gene to the prediction. SHAP is based on the concepts of game theory and can be applied to any machine learning model. The method uses Shapley values, which are derived from cooperative game theory, to fairly distribute the "payout" (i.e., the prediction) among the features. The formula for the Shapley value of gene $j$ is given by:

$$\phi_j = \sum_{S \subseteq \{1,...,p\} \setminus \{j\}} \frac{|S|!(p-|S|-1)!}{p!} (f(x_{S \cup \{j\}}) - f(x_S))$$

Here, $S$ represents a subset of genes excluding gene $j$, $p$ is the total number of genes, $|S|$ is the number of genes of S, $S \cup \{j\}$ represents a new subset formed by adding gene $j$ to the subset $S$, $f(x)$ is the prediction function of the model. Those genes with high Shapley values were considered as important genes.

We performed Kyoto Encyclopedia of Genes and Genomes (KEGG) enrichment analysis on the genes mined from the TransGeneSelector and Random Forest, using the STRING v11.5 database (<http://www.string-db.org). We focused on pathways with FDR < 0.05, considering them as significantly enriched.

**Network Analysis**

The modular regulatory network learning with per gene information (MERLIN) algorithm [55] was used to infer the regulatory network of the genes mined using TransGeneSelector or Random Forest in this study. The transcriptome data of the same dry seed and germinating seed datasets as well as another dataset unrelated to seed or seed germination of Arabidopsis thaliana were prepared (refer to Data description and processing). First, FPKM of the genes were transformed into transcripts per million reads (TPM), and the mean expression level of each gene was calculated. Then, the expression levels of genes were zero-mean transformed. Genes were retained for the MERLIN algorithm only if (1) their expression value varied by at least ± 1 from the mean in at least five samples and (2) they were in the list of genes obtained in the present study. All the genes were defined as both regulators and targets. A total of 10 sub-sets were created from the amended data matrix, each of which contained 50% of the samples selected randomly from the complete matrix. Data from each sub-set were used to infer a MERLIN interaction. In the final MERLIN network, each edge (which indicates the relation between two genes) appeared at least 6 times in the 10 sub-sets (confidence = 60%) were retained.

**Plant germination and RT-qPCR test**

We used the same collection of *Arabidopsis thaliana* col-0 mature seeds for both the germination experiment and the RT-qPCR quantitative analysis experiment. The seeds were stored at the Plant Development and Molecular Laboratory of Hunan Normal University, China.

We first selected surface-sterilized *Arabidopsis thaliana* seeds, sown on 9-cm plates containing solidified 0.5 MS medium (pH 5.9), stratified in the dark at 4°C for 2 d, and then exposed to different light strengths and durations. The light strengths were set from weak to strong, corresponding to aluminum foil wrapped full-black germination conditions, 100 μmol photons photosynthetic light, and 200 μmol photons photosynthetic light. For germination time, we set 0h (dry seeds), 12h, 24h, and 48h. The temperature for germination was set to 25 ℃.

Seed samples were ground to a fine powder in liquid nitrogen. The ground tissue was transferred to a pre-chilled 1.5-mL Eppendorf tube, and total RNA was isolated using the TRIzol R Reagent (Life Technologies, Carlsbad, CA, United States), according to the manufacturer's instructions. The RT-qPCR assay was conducted as described previously [56]. Forward primers used for genes are listed in Supplementary Table S1, and actin1 was used as reference. The expected size of the amplified fragments varied from 80 to 200 bp. Three technical replicates were performed for each sample. Statistical analysis was performed using Piko Real Software 2.0. After standardization of each gene expression, the heatmap drawing was performed using Python.